\documentclass[twocolumn, showpacs]{revtex4}
\usepackage{graphics}
\usepackage{amsmath}
\usepackage{mathrsfs}
\usepackage{amsfonts}
\usepackage{overpic}
\usepackage{hyperref}
\usepackage{rotating}
\newcommand{\ket}[1]{|#1\rangle}

\begin{document}
\title{Multiparticle Entanglement in the Lipkin-Meshkov-Glick Model}
\author{H. T. Cui }
\email{cuiht@aynu.edu.cn} \affiliation{Department of Physics, Anyang
Normal University, Anyang 455000, China}
\begin{abstract}
The multiparticle entanglement in the Lipkin-Meshkov-Glick model has
been discussed extensively in this paper. Measured by the global
entanglement and its generalization, our calculation shows that the
multiparticle entanglement can faithfully detect quantum phase
transitions. For an antiferromagnetic case the multiparticle
entanglement reaches the maximum at the transition point, whereas
for ferromagnetic coupling, two different behaviors of multiparticle
entanglement can be identified, dependent on the anisotropic
parameter in the coupling.
\end{abstract}
 \maketitle

\section{introduction}
The exploration of the connection between statistical mechanics and
quantum information has been extensive in recent years since the
work \cite{preskill}. Especially the research of entanglement in
many-body systems has contributed to the comprehensive crossover
between the two hot areas \cite{afov07}. Furthermore, the finding of
integer or fractional quantum Hall effect in two-dimensional
many-body systems imposes a challenge on the universal understanding
of phase transitions, since the traditional theory for phase
transition cannot incorporate these novel phenomena \cite{senthil}.
Recently the research of quantum entanglement in two-dimensional
many-body systems provides the clear characterization for different
quantum orders \cite{kitaev}. These facts suggest that quantum
entanglement would play a vital role in the understanding of
many-body effects.

Bipartite entanglement was first studied, and focused on the
connection to the criticality in spin-chain systems\cite{oaff02,
on02}. This interest comes from the fact that quantum phase
transition is related to the construction of the long-range
correlations in many-body systems. Hence it is a natural conjecture
that quantum entanglement, as a depiction of the non-local
correlation, could detect the appearance of long-range correlation.
Great progress has been made for the block entanglement in many-body
systems; the area law of block entanglement entropy has been
generally constructed by the conformal field theory. Furthermore the
violation of the area law has been identified as a reliable
detection of quantum phase transitions in one-dimensional
systems(see Ref. \cite{afov07} for a comprehensive review). However
the situation becomes complex for high-dimensional systems: the
violation of the area law in one-dimensional case when the system is
critical, does not seem to hold in higher dimensions\cite{afov07}.
Even for pairwise entanglement in many-body systems the results are
not satisfying . For example, the cutoff in the definition of
concurrence may lead to unphysical results when one focuses on the
connection of entanglement and phase transition in many-body systems
\cite{yang}.

The situation becomes more complex for multiparticle
entanglement(ME) because of the absence of unified measurement for
ME\cite{pv07}. However, it is a natural speculation that ME should
play a more fundamental role than a bipartite one for the
understanding of many-body effects with consideration of the
universal interaction in many-body systems. Recently the discussions
of ME in many-body systems have been given more attention because of
the availability of some special entangled states, e.g., cluster
states for one-way quantum computation\cite{br01}, $n$-party
Greenberger-Horne-Zeilinger(GHZ) state and $W$-state\cite{dvc00}.
Although great effort has been devoted to the measurement of ME, the
analytical or operational measurements exist only for some special
cases\cite{pv07}. The connection of quantum phase transition and ME
has also been discussed extensively in \cite{wei05,cffp07,
oliveira06}. However these discussions are mainly on the
one-dimensional spin-1/2 $XY$ model, and the difficulty of
calculating ME obstructs further exploration.

Recently, the global entanglement has been constructed for the
quantification of ME by Meyer and Wallach\cite{mw02}, which
possesses the virtues of the availability of analytical expression
and operability. Moreover global entanglement is measurable
experimentally since it is directly related to the mixedness of
single party\cite{brennen03}. Another important character of global
entanglement is the monotonicity under local operations and
classical communication (LOCC), if one notes that global
entanglement is intimately related to the linear
entropy\cite{horodecki07}. Consequently, Oliveira and his
collaborators improved this definition for measuring some special
entangled states, e.g. $n$-party $W$ state or GHZ
state\cite{oliveira06}. Moreover the connection of the generalized
global entanglement and quantum phase transition has also been
explored in the one-dimensional spin-1/2 $XY$ model, in which
entanglement reaches the maximum near to the transition
point\cite{oliveira06}.

It is interesting to note that the nearest neighbor coupling is
beneficial to the formation of ME in the one-dimensional spin-1/2
$XY$ model. Since the particle correlation is short-ranged in this
model\cite{on02}, one should note that the maximum of global
entanglement maybe come from the distribution of pairwise
entanglement\cite{cffp07, facchi}. Hence it is tempting to present a
discussion about ME  when the correlation is long range and the
coupling is beyond the nearest neighbor case. Fortunately the
Lipkin-Meshkov-Glick(LMG) model\cite{lmg} provides us the benchmark
for exploring this point since the collective interaction in this
model. Then it is expected that ME would play a critical role.

Recently the entanglement in the LMG model has been extensively
studied, such as concurrence \cite{vidal, dv04, vmd04},
one-tangle\cite{vpa04}, entanglement entropy\cite{uv, lp05,vdb07}
and generalized entanglement\cite{somma}. The concurrence in the LMG
model displays sensitivity to the appearance of quantum phase
transition\cite{vidal, vmd04}, except for some special
cases\cite{vmd04}. A possible explanation of this discrepancy is
that the trace operation performed in the calculation of concurrence
inevitably kills some correlations between spins\cite{vmd04}. With
respect to this point, Barnum and his collaborators constructed a
subsystem-independent measure of entanglement, based on a
distinguished subspace of observables for the system\cite{bkosv04}.
The named \emph{generalized entanglement} introduced by Barnum,
\textit{et.al.} has also been discussed in the LMG model, which
displayed the ability of detecting the phase
transitions\cite{somma}. Moreover the authors show the equivalency
between generalized entanglement and the global entanglement defined
by Meyer and Wallach\cite{mw02}. However, as shown in \cite{somma},
it is indispensable for the construction of the distinguished
subspace of observables to obtain the knowledge of the ground state
in many-body systems, that in most cases is very difficult. The
research of entanglement entropy in the LMG model shows that the
entropy was divergent under thermodynamic limit near the phase
transition point, and moreover shows a discontinuity at the critical
point for the isotropic coupling case\cite{vdb07}.

The generalized global entanglement (gGE), defined by Oliveira,
\textit{et. al.}\cite{oliveira06}, is a generalization of the global
entanglement (GE). With respect to the equivalence between Barnum's
generalized entanglement and GE, gGE provides a universal
characterization of ME in many-body systems. Hence, it is
interesting to present a comprehensive research of gGE and GE in the
LMG model. Our discussion also presents  detailed research for
antiferromagnetic coupling and some interesting results can be
obtained, which is rarely touched on in the previous works. I should
point out that the goal for this paper focuses on the connection
between ME, measured by gGE and GE respectively, and quantum phase
transition in the LMG model. For this purpose, the paper is
organized as following. In Sec.II the Hamiltonian is presented, and
ground states are determined analytically. The phase diagram will be
identified by introducing the proper parameter. In Sec.III the
analytical expressions for gGE and GE are presented. Based on these
formulas the multiparticle entanglements for ferro-magnetic and
antiferro-magnetic couplings are discussed respectively. The
conclusions and discussions are given in Sec. IV.

\section{Hamiltonian and ground state}
The LMG model describes a set of spin-half particles coupled to all
others with an interaction independent of the position and the
nature of the elements. The Hamiltonian can be written as
\begin{equation}\label{h}
H= - \frac{\lambda}{N}(S^2_x + \gamma S^2_y) - h_z S_z,
\end{equation}
in which $S_{\alpha}=\sum_{i=1}^{N}\sigma^i_{\alpha}/2 (\alpha=x, y,
z)$ and the $\sigma_{\alpha}$ is the Pauli operator, and $N$ is the
total particle number in this system. The prefactor $1/N$ is
essential to ensure the convergence of the free energy per spin in
the thermodynamic limit. Anti-ferromagnetic or ferromagnetic
interaction can be obtained dependent on $\lambda<0$ or
not($\lambda\neq0$). The Hamiltonian preserves the total spin and
does not couple the state having spin pointing in the direction
perpendicular to the field, i.e.
\begin{equation}\label{s}
[H, \textbf{S}^2]=0,   [H, \prod_{i=1}^{N}\sigma_z^i]=0.
\end{equation}
For isotropic coupling $\gamma=1$, $[H, S_z]=0$ and the spectrum of
Eq.\eqref{h} can be determined exactly. However for $\gamma\neq 1$,
the spectrum can be determined in principle by Bethe-type
equations\cite{plo} and the analytical expressions are difficult to
obtain.

A distinguished character of Eq. \eqref{h} is the collective
interaction, which is the same for any particle and independent of
the space configuration of the system. Because of long-range
correlation between particles, the mean-field analysis is adaptive
for this model\cite{botet}. The research of phase transition in the
LMG model has shown that there is a second-order transition at
$h=h_z/|\lambda|=1$ for the ferromagnetic case and a first-order one
at $h=0$ for the antiferromagnetic case\cite{botet, vidal}.

A proper parameter for characterizing the phase diagram is the total
spin in the direction $z$ for the ground state. For ferromagnetic
coupling, one has
\begin{eqnarray}
1-2\langle S_z\rangle/N=\begin{cases}0, &h>1\\1-h, &h\in[0,
1),\end{cases}
\end{eqnarray}
which corresponds to the disorder-order transition, and obviously
the point $h=1$ is a second-order phase transition point. This phase
transition could be attributed to the disappearance of the energy
gap; at the symmetric phase $h>1$ the energy gap above the ground
state is finite, whereas at the broken phase $0\leq h<1$ the energy
gap vanishes under thermodynamic limit\cite{dv04}. For
antiferromagnetic coupling,
\begin{eqnarray} 2\langle
S_z\rangle/N=\begin{cases}1, &h>0\\-1, &h<0;\end{cases}
\end{eqnarray}
Obviously there is a first-order phase transition at the point
$h=0$. For this case the energy gap above ground state vanishes only
at the transition point $h=0$ under thermodynamic limit,  and no
level crossing happens when $h\neq0$\cite{vmd04}.

The ground state for $\gamma\neq1$ can be determined analytically
with the help of Holstein-Primakoff(HP) transformation and
low-energy approximation\cite{dv04}. In Ref. \cite{cui06}, the
ground state has been obtained with the consideration of the finite
number effect. The general expression reads
\begin{eqnarray}
\label{g}
\ket{g}&=&\frac{1}{c}\sum_{n=0}^{[N/2]}(-1)^n\sqrt{\frac{(2n-1)!!}{2n!!}}\tanh^nx\ket{2n}\nonumber\\
c^2&=&\sum_{n=0}^{[N/2]}(-1)^n\frac{(2n-1)!!}{2n!!}\tanh^{2n}x
\end{eqnarray}
in which $\ket{2n}$ is the Fock state of the boson operator
introduced by Holstein-Primakoff transformation and $[N/2]$ denotes
the integer part not more than $N/2$. One should note that the
determination of the ground state Eq. \eqref{g} is based on HP
transformation, which preserves the symmetry Eq. \eqref{s}, and the
following discussion is heavily based on this ground state.
Dependent on the style of interaction, $\tanh \_ x$ has different
expressions. For ferromagnetic case $\lambda>0$, it satisfies the
relation\cite{vidal}
\begin{eqnarray}\label{ft}
\label{ferro} \tanh 2x=\begin{cases}-\frac{1-\gamma}{2h-1-\gamma},&
h>1\\-\frac{h^2-\gamma}{2-h^2-\gamma}, & 0\leq h<1\end{cases}.
\end{eqnarray}
For antiferromagnetic coupling $\lambda<0$, it is determined by
\begin{eqnarray}\label{aft}
\tanh\_2x=\frac{1-\gamma}{1+\gamma+2|h|}.
\end{eqnarray}

For isotropic case $\gamma=1$, the calculation is exact. The ground
state can be formulated generally as $\ket{g}=\ket{S=\frac{N}{2},
S_z=M}$. For ferromagnetic coupling,
\begin{eqnarray}\label{gf}
M=\begin{cases}I[h N/2],& 0\leq h<1\\ \frac{N}{2}, & h\geq1.
\end{cases}
\end{eqnarray}
in which $I[n]$ expresses the integer not more than $n$. For
antiferromagnetic coupling ,
\begin{eqnarray}\label{ga}
M=\begin{cases}\frac{N}{2},& h>0\\ -\frac{N}{2}, & h<0.
\end{cases}
\end{eqnarray}

\section{Multiparticle Entanglement}
Recently, Meyer and Wallach have constructed the global entanglement
for measuring ME in spin systems. The main procedure is to first
measure the entanglement between any party and the others, and then
calculate the average of all possible bipartition\cite{mw02}.
Although the criticism that it is not a genuine ME measure because
of the intimate connection to bipartite entanglement
\cite{horodecki07}, it has been proven that the global entanglement
is operational and more importantly,  monotonic under LOCC. A
simplified expressions for global entanglement is provided by
Brennen \cite{brennen03}
\begin{equation}
Q(\ket{\phi})=2(1-\frac{1}{N}\sum_{k=0}^{N-1}\text{Tr}[\rho_k^2]).
\end{equation}
For the LMG model, one can obtain
$Q(\ket{g})=2(1-\text{Tr}\rho_1^2)$, in which $\rho_1$ stands for
the single-particle reduced density operator. Furthermore, Oliveira
and his collaborators have improved this definition in order that it
can measure some special entangle states, e.g. $\otimes_n
\ket{EPR}_n$ or $n$-party GHZ state. The main procedure is to
measure the entanglement between any two parties and the others, and
then average all possible bipartition\cite{oliveira06}. In the LMG
model, for the symmetry of particle permutation, the generalized
global entanglement can be written as
\begin{equation}
E_g=\frac{4}{3}(1- \text{Tr}[\rho^2_{2}]).
\end{equation}
in which $\rho_{2}$ denotes the reduced density operator for any two
particles.  $\rho_1,\rho_{2}$ can be determined through the
correlation functions\cite{wm02}
\begin{eqnarray}\label{c}
\langle\sigma_{\alpha}\rangle&=&\frac{2}{N}\langle
S_{\alpha}\rangle,
\nonumber\\
\langle\sigma_{1\alpha}\sigma_{2\alpha}\rangle&=&\frac{4\langle
S^2_{\alpha}\rangle-N}{N(N-1)}\nonumber\\
\langle\sigma_{1\alpha}\sigma_{2\beta}\rangle&=&\frac{2\langle
[S_{\alpha}, S_{\beta}]_+\rangle-N}{N(N-1)}(\alpha\neq\beta)
\end{eqnarray}
in which $\alpha, \beta = x, y, z$.

With respect to  the symmetry Eq. \eqref{s} and the ground state Eq.
\eqref{g}, one can obtain GE and gGE respectively
\begin{eqnarray}\label{me}
&Q(\ket{g})=1 - \langle\sigma_z\rangle^2\nonumber \\ &E_g=1 -
\frac{1}{3}(2\langle\sigma_z\rangle^2+\langle\sigma_{1x}\sigma_{2x}\rangle^2
+\langle\sigma_{1y}\sigma_{2y}\rangle^2+\langle\sigma_{1z}\sigma_{2z}\rangle^2).
\end{eqnarray}
Based on Eqs. \eqref{g} and \eqref{c}, ME in the  LMG model can be
decided analytically, and some interesting properties can be found.
The discussion below is divided into two cases: one focuses on the
anisotropic coupling, and the other is for isotropic coupling for
which the exact results can be obtained.

\subsection{anisotropic coupling}
The analytical results can be obtained under large $N$ limit with
the hypothesis that the excitation would only happen for the low
energy states\cite{vidal}. Based on Eqs. \eqref{g} and \eqref{c},
one obtains
\begin{widetext}
\begin{eqnarray}
\langle\sigma_z\rangle&=&1-\frac{4}{Nc^2}\sum_{n=0}^{[N/2]}nc_{2n}^2\nonumber\\
\langle\sigma_{1x}\sigma_{2x}\rangle&=&\frac{2}{N(N-1)c^2}\sum_{n=0}^{[N/2]}
[\sqrt{(N-2n+2)(N-2n+1)2n(2n-1)}c_{2n-2}c_{2n}+2n(N-2n)c^2_{2n}]\nonumber\\
\langle\sigma_{1y}\sigma_{2y}\rangle&=&\frac{-
2}{N(N-1)c^2}\sum_{n=0}^{[N/2]}
[\sqrt{(N-2n+2)(N-2n+1)2n(2n-1)}c_{2n-2}c_{2n}-2n(N-2n)c^2_{2n}]\nonumber\\
\langle\sigma_{1z}\sigma_{2z}\rangle&=&1-\frac{4}{N(N-1)c^2}\sum_{n=0}^{[N/2]}2n(n-2n)c^2_{2n}
\end{eqnarray}
\end{widetext}
in which $c_{2n}=(-1)^n\sqrt{\frac{(2n-1)!!}{2n!!}}\tanh^n\_x$ and
$\tanh x$ is decided by Eqs.\eqref{ft} and \eqref{aft}. From Eq.
\eqref{me}, ME can be determined analytically.

\begin{figure}[tb]
\includegraphics{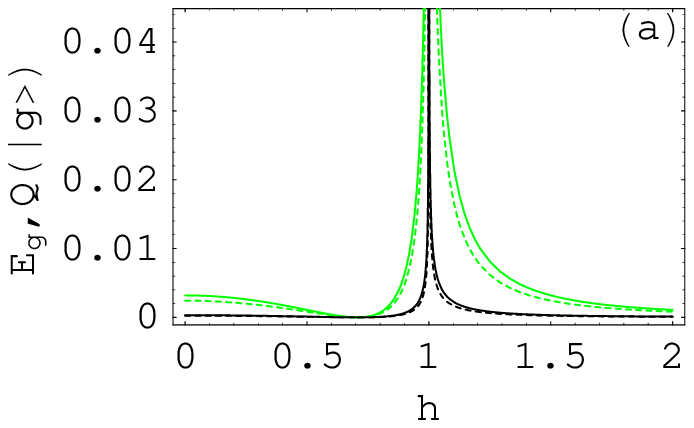}
\includegraphics{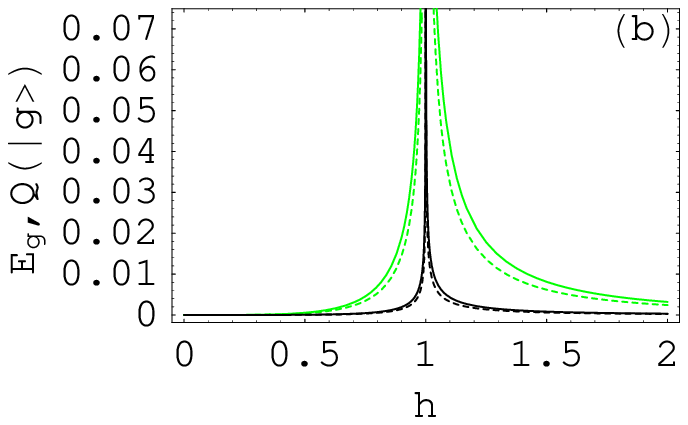}
\caption{\label{f1}(Color online) The multiparticle entanglement for
ferromagnetic coupling, measured by gGE (denoted by $E_g$ and solid
lines) and GE (denoted by $Q(\ket{g})$ and dashed lines) vs the
rescaled magnetic field $h$. We have chosen $\gamma=0.5$(a) and
$\gamma=0$(b) for this plot. The green and black solid lines
correspond to $N=50, 500$ respectively. }
\end{figure}

\begin{figure}[tb]
\includegraphics[bbllx=14, bblly=15, bburx=265, bbury=218, width=7cm]{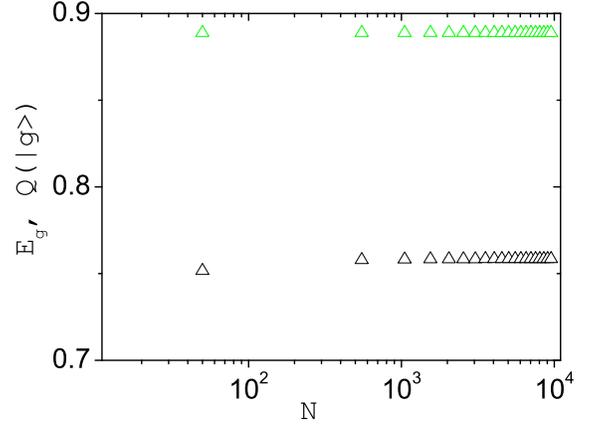}
\caption{\label{f2} (Color online) $E_g$ (black triangles) and
$Q(\ket{g})$ (green triangles) for ferromagnetic coupling vs. the
particle number $N$ at the phase transition point $h=1$. One should
note that $\tanh 2x$ for any $\gamma\in[0, 1)$ is identical in this
case. }
\end{figure}

\emph{-Ferromagnetic case-} gGE and GE have both been plotted for
different $\gamma$ in Fig.\ref{f1}. It is obvious that ME reaches
the maximum closed to phase transition point $h=1$ and the slope of
curves tends to be infinite. In  recent papers\cite{oliveira06}, the
authors have shown that the singularity of gGE is directly connected
to the degeneracy of the ground-state energy at the phase transition
point. Since the energy gap above ground state vanishes at critical
point $h=1$, the singularity of GE and gGE can be attributed to the
degeneracy of ground-state energy, and could be used as a reliable
detector for the phase transition in this case. Furthermore the
finite-size scaling at the phase transition point $h=1$ displays the
non-sensitivity both of gGE and GE to the particle number $N$, as
shown in Fig.\ref{f2}.

\begin{figure}[tb]
\includegraphics{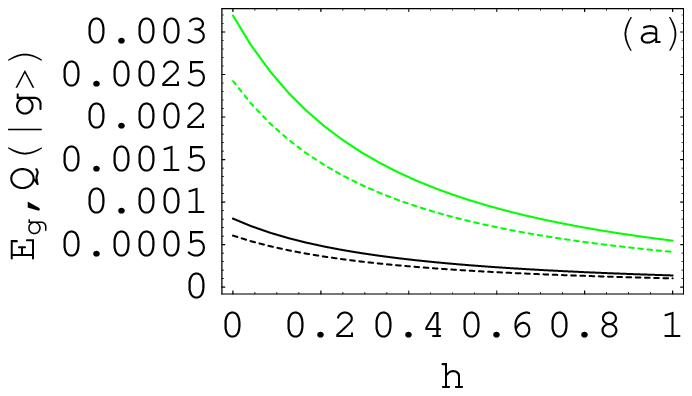}
\\[0.5cm]
\includegraphics{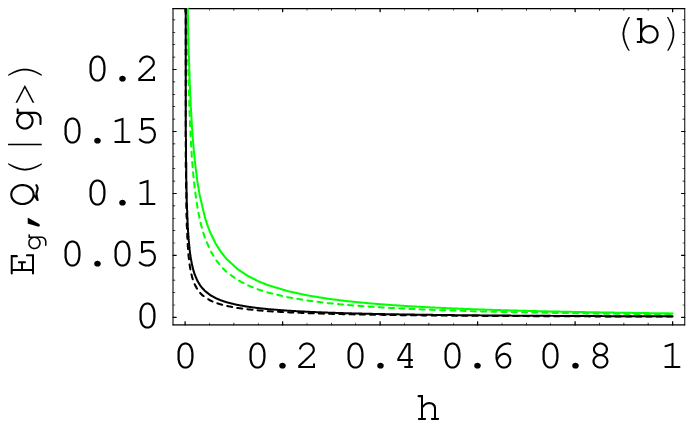}
\caption{\label{f3}(Color online) $E_g$ (solid lines) and
$Q(\ket{g})$ (dashed lines) for antiferromagnetic coupling vs. the
rescaled magnetic field $h$. The parameter $\gamma=0.5$(a) and
$\gamma=0$(b) have been chosen for this plot. Since the system is
invariant under the changing of $h\leftrightarrow -h$, these
plottings are only for $h\ge0$ with N=50 (green lines) and N=200
(black lines) respectively.}
\end{figure}

\begin{figure}[tb]
\includegraphics[width=8cm]{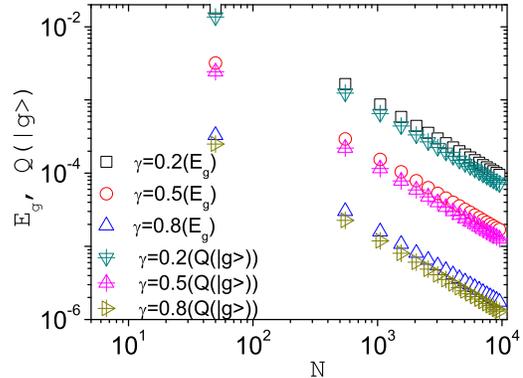}
\caption{\label{f4} (Color online) $E_g$ and $Q(\ket{g})$ for
antiferromagnetic coupling vs. the particle number $N$ at
first-order phase transition point $h=0$. We have choosen three
representative values of $\gamma$ for this plot.}
\end{figure}

\emph{-Antiferromagnetic case-} The situation is intricate. We have
plotted GE and gGE vs. $h$ by choosing $\gamma=0$ (a) and
$\gamma=1/2$ (b) respectively,  in Fig.\ref{f3}. Since there is a
first-order quantum phase transition, the figures show that GE and
gGE both are maximum at transition point $h=0$. However a further
calculation shows two different behaviors for ME; For
$\gamma\in(0,1)$ GE and gGE show a cusp at phase transition point
$h=0$, which means that the first derivation of gGE and GE with
respect to $h$ is discontinued but finite at $h=0$. Since the
degeneracy of ground-state energy happens only at $h=0$, this
discontinuity of gGE and GE is attributed to the degeneracy of
ground-state energy, whereas for $\gamma=0$, the figure shows that
gGE and GE both have a drastic increase closed to the phase
transition point, and the first derivations of gGE and GE with $h$
tend to be divergent. Furthermore our calculation shows that with
the increasing of particle number, gGE and GE decrease for
$\gamma\in(0,1)$ at the transition point, as shown in Fig. \ref{f4}.
Moreover, GE and gGE have similar behaviors and the difference
between them is slight.

\subsection{isotropic coupling}
When $\gamma=1$, the exact results can be obtained. With respect to
Eqs. \eqref{gf}, \eqref{ga} and \eqref{c}, one has in this case
\begin{eqnarray}\label{me2}
E_g&=&1-
\frac{8M^2}{3N^2}-\frac{2(4M^2-N)^2+(4M^2-N^2)^2}{6N^2(N-1)^2}\nonumber\\
Q(\ket{g})&=&1- (\frac{2M}{N})^2.
\end{eqnarray}

\begin{figure}[tb]
\includegraphics{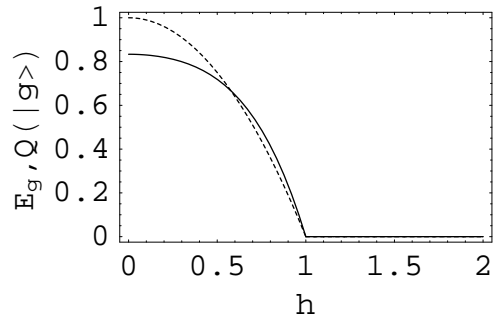}
\caption{\label{f5} $E_g$ (solid line) and $Q(\ket{g})$ (dashed
line) vs $h$ for isotropic ferromagnetic coupling with $N\rightarrow
\infty$.}
\end{figure}

\emph{-Ferromagnetic case-}  With respect to Eqs. \eqref{gf} and
\eqref{me2}, both gGE and GE are zero for $h\ge1$, independent of
the particle number $N$ since the ground state is a direct-product
state of $N$ particles with the same spin orientation from Eq.
\eqref{gf}. As shown in Fig. \ref{f5}, gGE and GE both are continued
under thermodynamic limit at the phase transition $h=1$. This
behavior is different from the conclusion made in Refs. \cite{lp05,
vdb07, dv04}, in which entanglement entropy and concurrence both are
discontinued at $h=1$ under thermodynamic limit. The main reason for
this discrepancy is stated below.  One notes that Eqs.\eqref{me2}
are a function of $M/N$. It is obvious from Eq. \eqref{gf} that
$M/N$ is continuous at the transition point $h=1$ under
thermodynamic limit. Hence, it is not strange that gGE and GE are
also continuous. Comparably the concurrence in \cite{dv04} is
redefined by adding a prefactor $N-1$ to keep finite under
thermodynamic limit. Similarly for the calculation of entanglement
entropy, all possible bipartition has to be considered in the
calculation to keep the entropy finite under thermodynamic
limit\cite{lp05}, which I point out plays the same function of the
prefactor $N-1$ for the calculation of concurrence. Under the
thermodynamic limit, the prefactor would play a nontrivial role.
Since the calculation for entanglement has been implemented
respectively in different regions because of Eq. \eqref{gf}, the
difference, which should disappear under thermodynamic limit, may
become finite because of this prefactor. While, since our definition
of ground state Eq.\eqref{g} has naturally considered the
finite-number effect and GE and gGE are the functions of the
correlations, one does not need this prefactor to keep the
measurements of entanglement finite under thermodynamic limit.
Together with respect that the equivalency between generalized
entanglement and GE have been proved\cite{somma}, the measure gGE
may also has the great virtue of independence on the concept of
subsystem.

\emph{-Antiferromagnetic case-} With respect to Eqs. \eqref{ga} and
\eqref{me2}, gGE and GE both vanish independently on $N$. Since the
states for $M=\pm N/2$ are degenerate at phase transition point
$h=0$, the ground state is undoubtedly the superposition of states
$\ket{N/2, N/2}$ and $\ket{N/2, -N/2}$, written on the basis of
$\{S^2, S_z\}$, with equal weight,
\begin{equation}
\ket{g}=\frac{1}{\sqrt{2}}(\ket{N/2, N/2}+\ket{N/2, -N/2}).
\end{equation}
Then in this case, a genuine maximally multiparticle entangled
state, so called n-party GHZ state\cite{dvc00}, can be obtained at
the point $h=0$ for a finite particle number, and under
thermodynamic limit, it corresponds to the celebrated (
Schr\"odinger cat ) macroscopic quantum superposition state.
Obviously, the measurement of entanglement is discontinued at $h=0$,
where a first-order phase transition happens under thermodynamic
limit\cite{footnote1}.

\section{discussions and conclusions}
Some comments and discussions should be presented. In this paper an
extensive discussion of multiparticle entanglement, measured by GE
\cite{mw02} and gGE \cite{oliveira06}, is presented in the
Lipkin-Meshkov-Glick model. Our discussion focuses on two different
situations: for ferromagnetic coupling $\lambda>0$, when the
anisotropic parameter $0<\gamma<1$ gGE and GE both reach the maximum
at the second-order phase transition point $h=1$, as shown in Figs.
\ref{f1}. Moreover they are nonsensitive to the variation of
particle number $N$, as shown in Fig.\ref{f2}. Whereas for the
isotropic case $\gamma=1$ gGE and GE are zero at the phase
transition point, shown in Fig. \ref{f5}.

Another important situation is the appearance of antiferromagnetic
coupling $\lambda<0$, for which there is a first-order phase
transition at transition point $h=0$. gGE and GE both are
calculated, as shown in Fig. \ref{f3}. It is interesting that some
different behaviors can be found in this case; one is that gGE and
GE have a cusp at phase transition point when $\gamma\in(0,1)$,
which means that the first derivation of gGE and GE with external
magnetic field is discontinued but finite at $h=0$, shown in Fig.
\ref{f3}(a). Another case happens when $\gamma=0$, in which both gGE
and GE have a drastic changing closed to $h=0$, shown in Fig.
\ref{f3}(b). Moreover our calculations show that gGE and GE for
$\gamma\in(0,1)$ decrease with the increment of $N$ as shown in
Fig.\ref{f4}, whereas for $\gamma=0$ they are non-sensitive to the
particle number $N$. It is more interesting for the isotropic case
that the entanglement is vanishing for $h\neq0$ and has a sudden
changing at $h=0$, where a genuine maximally multiparticle
entanglement state, $n$-party GHZ state for finite particle number,
can be obtained. With connection of a scheme of the realization of
the LMG model in optical cavity QED\cite{mp07}, this result provides
a powerful method to create ME experimentally.

It was naturally expected that ME should be maximum at the phase
transition point since the correlation between particles would be
long-range because of the appearance of critical quantum fluctuation
at the phase transition point. However an exceptional case appears
in our discussion, which happens for ferromagnetic and isotropic
coupling. In my own opinion, a reason for the difficulty in
constructing the connection between entanglement and quantum phase
transition is that the up-to-date measurements for entanglement are
generally a nonlinear function of correlation functions in many-body
systems. Hence the singularity of correlation functions may be
canceled\cite{footnote2}. As a concrete illustration one notes that
for antiferromagnetic coupling, $M$ has a discontinued change at
$h=0$ for $\gamma=1$, as shown in Eq.\eqref{ga}. However from
Eq.\eqref{me2} it is obvious that the discontinuity of $M$ for $h>0$
and $h<0$ has no effect on ME since gGE and GE are the functions of
$M^2$.

Regardless of this defect, some interesting information can be
obtained from the research of ME. An interesting speculation from
our discussion is that the different finite-size scales may show the
different state structures for the entanglement at the phase
transition point. As shown previously in Ref. \cite{oliveira06},
with the increment of particle number the measures for entanglement
for some states are decreasing, whereas for other states tend to be
steady values. Since the increment of particle number, or more
generally the degree of freedom, means the stronger correlation
between the particles in many-body systems, one can conclude that
some entangled states are immune to the effect imposed by the
increment of correlation between particles, whereas others are
sensitive to the changing of correlation. With respect to the
pursuit of decoherence-free space\cite{dfs}, our discussion may
provide some useful information.

With connection to the researches of the bipartite entanglement in
many-body systems, one could note that it is difficult to construct
a universal classification of phase transition based on the
entanglement and its derivation. The main obstacle, in my own
opinion, is the absence of physical definition of entanglement,
i.e., how and what to define an "entanglement operator". Since
quantum entanglement is an important physical resource and can be
measured experimentally, I believe in the existence of this
operator. Fortunately a few works have attributed to this
direction\cite{somma, cv06}. I hope our discussion will help in the
understanding of entanglement in many-body systems.

\emph{Acknowledgement} The author acknowledges the support of
Special Foundation of Theoretical Physics of NSF in China, Grant No.
10747159.


\begin{thebibliography}{99}
\bibitem{footnote1} Eq. \ref{me2} is obviously failed to measure the
n-party GHZ state since its dependence on $M^2$ and the
discontinuity of $M$ has no effect on ME.
\bibitem{footnote2}The continuity of function $f(x)$ is defined as
\begin{equation*}
\lim_{x\rightarrow x_0}f(x)-f(x_0)=0.
\end{equation*}
First suppose that $f(x)$ is a linear function of the subfunctions
$g_k(x)(k=1, 2, \dots, n)$, i.e. $f(x)=\sum_{k=1}^n c_k g_k(x)$ in
which $c_k$ is the coefficient. Without lost of generality any
$g_k(x)$ is supposed not to be the linear combination of the other
$g_k(x)$. Then the continuity of $f(x)$ requires
\begin{eqnarray*}
&\sum_{k=0}^n&c_k \lim_{x\rightarrow x_0}g_k(x)-g_k(x_0)=0, \\
&\Rightarrow& \lim_{x\rightarrow x_0}g_k(x)-g_k(x_0)=0
\end{eqnarray*}
The conclusion above means that the continuity of $g_k(x)$ is
sufficient and necessary condition for the continuity of $f(x)$.

With respect of nonlinear case, let consider the function $f[g(x)]$
for simplicity. From the definition of the continuity and suppose
$\delta x=x-x_0$, it requires
\begin{eqnarray*}
&\lim_{x\rightarrow x_0}&f[g(x)]-f[g(x_0)]=0\\
&\Rightarrow& \lim_{\delta x\rightarrow 0}\frac{\partial
f[g(x)]}{\partial x}|_{x=x_0}\delta x=0\\
&\Rightarrow& \lim_{\delta x\rightarrow 0}\frac{\partial
f[g(x)]}{\partial g(x)}|_{x=x_0}[g(x)-g(x_0)]=0.
\end{eqnarray*}
The prefactor $\frac{\partial f[g(x)]}{\partial g(x)}|_{x=x_0}$ has
a nontrivial effect; Even if $\lim_{\delta x\rightarrow
0}g(x)-g(x_0)\neq 0$, the condition of continuity may still be
satisfied only if this prefactor vanishes and $\lim_{\delta
x\rightarrow 0}g(x)-g(x_0)$ keeps finite.
\end{thebibliography}
\end{document}